\documentstyle[pre,aps,multicol,psfig,rotate]{revtex}
\begin{document}
\draft
\title{Critical point shift in a fluid confined between opposing walls}
\author{Enrico Carlon}
\address{Institute for Theoretical Physics, Katholieke Universiteit 
Leuven, Celestijnenlaan 200D, B-3001 Leuven, Belgium}
\author{Andrzej Drzewi\'nski}
\address{Institute for Low Temperature and Structure Research, 
Polish Academy of Sciences, P. O. Box 937, 50-950 Wroc\l aw 2, Poland}
\date{Published in \pre {\bf 57}, 2626 (1998)}
\maketitle
\begin{abstract}
The properties of a fluid, or Ising magnet, confined in a $L \times \infty$
geometry with opposing surface fields at the walls is studied by density 
matrix renormalization techniques. In particular we focus on the effect of 
gravity on the system, which is modeled by a bulk field whose strength varies 
linearly with the distance from the walls. It is well known that in the absence 
of gravity phase coexistence is restricted to temperatures below the wetting 
temperature. We find that gravity restores phase coexistence up to the bulk 
critical temperature, in agreement with previous mean field results. 
A detailed study of the scaling to the critical point, as $L \to \infty$, 
is performed. The temperature shift scales as $1/L^{y_T}$, while the 
gravitational constant scales as $1/L^{1+y_H}$, with $y_T$ and $y_H$ the 
bulk thermal and magnetic exponents respectively.
For weak surface fields and $L$ not too large, we also observe a regime 
where the gravitational constant scales as $1/L^{1+y_H - \Delta_1 y_T}$
($\Delta_1$ is the surface gap exponent) with a crossover, for sufficiently 
large $L$, to a scaling of type $1/L^{1+y_H}$.
\end{abstract}

\begin{multicols}{2} \narrowtext
\section{Introduction}

The Ising model has played a central role in the theory of critical 
phenomena. It is simple enough that it can be studied in great detail 
(although exact solutions are restricted  only to a few cases in two 
dimensions), and it can be used to model many interesting physical 
situations. The two phases with opposite magnetization, which coexist 
in an infinite system for temperatures lower than the bulk critical 
temperature $T_c$, can be thought as the two coexisting phases (liquid 
and vapor) of a simple fluid.
Besides bulk critical phenomena, which occur in a system infinitely 
extended in all directions, much interest has been focused on the 
study of the critical behavior of confined systems, which are of 
finite extensions in one or more directions.

A few years ago Parry and Evans \cite{parry,parryA}, using a mean field 
Ginzburg-Landau approach, investigated the phase diagram of a $d$-dimensional 
Ising model in a $L \times \infty^{d-1}$ geometry, i.e. confined between 
two walls separated by a finite distance $L$. 
They considered opposing walls, where one wall favors the ``liquid" and 
the other the ``vapor" phase; in the Ising language this corresponds to 
introducing surface magnetic fields $h_1$ and $h_2$ with opposite sign 
($h_1 h_2 < 0$). 
It was found \cite{parry} that two phase coexistence is restricted to 
temperatures below the wetting temperature $T_w$; Brochard and de Gennes
had come to a similar conclusion as well \cite{broch}.
These results were later confirmed by extensive Monte Carlo simulations, 
both in two and three dimensions \cite{albano,albano2,binder2}.
The two dimensional Ising model with opposing surface fields was recently
solved exactly by Macio\l ek and Stecki \cite{maciolek} using transfer 
matrices and pfaffian techniques.
The surprising characteristic of Parry and Evans' results \cite{parry,parryA}
is that one does not seem to recover information about the bulk critical point 
when the limit $L \to \infty$ is taken: for all values of $L$ two phase 
coexistence is restricted to $T < T_w$.

In trying to clarify the remarkable properties of this system, Rogiers and 
Indekeu \cite{jos} introduced an extra bulk field which varies linearly 
with the distance from the walls and which plays the role of gravity in the 
fluid. They found, in a mean field analysis, that the competing effect of 
opposing surface fields and gravity restores phase coexistence up to the bulk
critical temperature.

In the present paper we study the model of Rogiers and Indekeu at the lower 
critical dimension ($d=2$) and beyond the mean field approximation to test the 
validity of their conclusion when thermal fluctuations are properly taken into 
account.
Our results are based on a density matrix renormalization group (DMRG)
calculation and essentially confirm the mean field scenario. We also 
analyze the exponents describing the critical point shift along the 
thermal and gravitational field direction and compare our findings with 
some predictions due to scaling analysis \cite{jos,fisher,nakanishi,vanleeuwen}.

A brief report of this work has been presented in Ref. \cite{prl}; 
here we give a full account of our work and present a series of results 
concerning in particular the analysis of the finite size scaling of the 
gravitational constant for low surface fields, where interesting 
crossover phenomena take place. 
This paper is organized as follows. In Sec.\ref{sec:mod} we introduce 
the model and in Sec.\ref{sec:DMRG} briefly review the DMRG technique.
In Sec.\ref{sec:Tw} we compare DMRG results with exact ones in the absence of
a gravitational field. In Sec.\ref{sec:phd} we present the phase diagram 
as derived from DMRG calculations and discuss its main features. In
Sec.\ref{sec:pro} we show some magnetization profiles and compare the 
DMRG results with those obtained by capillary wave theory.
In Sec. \ref{sec:con} we present some conclusions.

\section{Model}
\label{sec:mod}

We consider a ferromagnetic Ising model in a $L \times \infty$ strip
with the following Hamiltonian:
\begin{eqnarray}
H = -J \sum_{i,j} s_{i,j} s_{i+1,j} -J \sum_{i,j} s_{i,j} s_{i,j+1} + 
h_1 \sum_j s_{1,j} 
\nonumber \\
- h_1 \sum_j s_{L,j} + 
g \sum_j \sum_{i=1}^{L} (2 i - 1 - L) s_{i,j},
\label{ham}
\end{eqnarray}
where $J > 0$, $s_{i,j}=\pm 1$, $-\infty < j < \infty$ and $1 \leq i \leq L$. 
For simplicity we restrict ourselves to the case of surface fields of equal 
magnitude $|h_1|$. The last term of the right-hand side of Eq. (\ref{ham}) 
is a bulk field varying linearly with the distance from the walls, which, 
for convenience, is chosen antisymmetric with respect to the center of 
the strip.
This field models the effect of gravity on a fluid and $g$ plays the role 
of the gravitational constant. 

In the limit $g \to 0$ one recovers the models studied by Parry and Evans 
\cite{parry}; in two dimensions the wetting temperature $T_w$ is known 
exactly and satisfies the equation \cite{abraham}
\begin{eqnarray}
e^{a} \left[ \cosh (a) - \cosh (b) \right] = \sinh (a),
\label{twexact}
\end{eqnarray}
where $a=2J/T_w$ and $b=2h_1/T_w$. 
For $h_1 = 0$ the wetting temperature $T_w$ is equal to the bulk critical
temperature $T_c = 2 J/\ln(1+\sqrt{2})$. It decreases monotonically with 
$h_1$ and vanishes for $h_1 = J$.

In the rest of the paper we will be primarily interested in the competing 
effect of surface fields and gravity; it is clear from the Hamiltonian 
(\ref{ham}) that these occur when $h_1$ and $g$ have the same sign.

\section{DMRG method}
\label{sec:DMRG}

The density matrix renormalization group was introduced by White 
\cite{whitePRL,whitePRB} to study the ground state properties of quantum 
spin chains. The method is very accurate and it has been successfully 
applied to many one dimensional quantum problems (for a review see 
Ref. \cite{gehring}).

Exploiting the relation between a $d$-dimensional quantum system and a 
$d+1$-classical system, Nishino \cite{nishino1} was able to extend the 
DMRG to two-dimensional classical systems. In this case White's algorithm 
is applied to the construction of effective transfer matrices of large
systems.

In a transfer matrix approach the partition function of a system
defined on a $L \times \infty$ strip is equal to the largest 
eigenvalue of the so-called transfer matrix $T_L$ \cite{baxter}. 
Numerical calculations are restricted to strips of small widths 
(typically $L < 20$ for an Ising model) since the dimension of 
$T_L$ grows exponentially with the strip width.

In a DMRG calculation one starts from a transfer matrix of a small system
that can be handled exactly. 
Using the information about the thermodynamics of this system one generates
an effective transfer matrix of a larger system. The strip width grows at 
each DMRG iteration and the spin space is very efficiently truncated to
keep the dimensionality of the effective transfer matrix controlled.

\begin{figure}[b]
\centerline{
\rotate[r]{\psfig{file=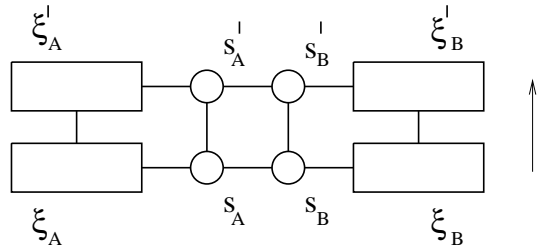,height=7cm}}}
\vskip 0.2truecm
\caption{Schematic view of a transfer matrix element of a strip of width $L$ 
generated by DMRG. $\xi$ and $s$ label block and spin variables respectively, 
with $\xi = 1,2 \ldots m$ and $s = \pm 1$. The total dimension of the matrix 
is $4 m^2 \times 4 m^2$. The arrow denotes the transfer direction.}
\label{FIG01}
\end{figure}

Figure \ref{FIG01} shows schematically a transfer matrix element generated by 
the DMRG algorithm. The matrix consists of block and spin variables indicated 
in the figure by rectangles and circles respectively. A block, whose states 
are labeled by a variable $\xi$ which can take $m$ possible values only, 
describes approximately a collection of spins. 

Obviously by allowing larger $m$ one obtains more accurate numerical 
results; in a typical DMRG calculation the accuracy grows very rapidly 
with $m$ \cite{whitePRL}. In the present case we found that for strips 
up to $L = 100$ a value of $m = 40$ is sufficient to guarantee a very 
high accuracy of the numerical results.
In our calculations we have used the finite system method described by
White in Ref.\cite{whitePRB}, a version of the DMRG algorithm designed to 
accurately study finite size systems. For more details we refer readers to the 
existing literature (see Refs.\cite{whitePRL,whitePRB,gehring,nishino1}).

Note that we use open boundary conditions: the blocks on the left and 
right side of Fig. \ref{FIG01} are not coupled together. Although it is 
possible to implement the DMRG method with periodic boundary conditions 
it turns out that the accuracy is lower than in the open boundary 
condition case \cite{whitePRB}. 
Therefore the method is best suited to study properties of two dimensional
classical systems in contact with walls or with free surfaces.

\section{Zero-gravity case}
\label{sec:Tw}

In the limit $g \to 0$ the model described by the Hamiltonian (\ref{ham}) has 
been solved exactly \cite{maciolek}. In this section we summarize briefly some 
known facts and present results obtained from DMRG calculations. 

\begin{figure}[b]
\centerline{
\rotate[r]{\psfig{file=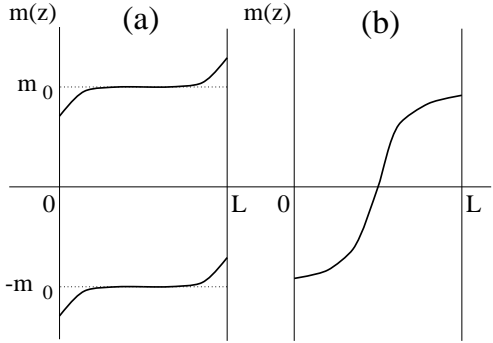,height=6.5cm}}}
\vskip 0.2truecm
\caption{Magnetization profiles for the model in absence of gravity 
in the two phase coexistence region for $0 \leq T < T_d(L)$ (a) and 
in the single phase region for $T_d(L) \leq T < T_c$ (b). 
$m_0$ denotes the bulk magnetization of the Ising model.}
\label{FIG02}
\end{figure}

For temperatures below a temperature $T_d(L)$, which Parry and Evans 
\cite{parry} interpreted as the critical temperature of the confined 
system, two phases coexist. Typical magnetization profiles are shown 
in Fig. \ref{FIG02}(a). For $T_d(L) \leq T < T_c$ the system is in a 
single interface-like state as depicted in Fig. 2(b). 
For $L \to \infty$ $T_d(L)$ scales as \cite{parry}:
\begin{eqnarray}
T_d(L) - T_w \sim L^{-1/\beta_s}.
\label{wet}
\end{eqnarray}
Here $\beta_s$ is the exponent describing the divergence of the thickness 
of the wetting layer for a semi-infinite system with a surface field $h_1$: 
$l \sim (T_w - T)^{-\beta_s}$.

Swift {\em et al.} \cite{swift}, who analyzed the same system from a different 
angle, interpreted $T_d (L)$ rather as a shifted wetting temperature, in 
contrast with the point of view of Ref.\cite{parry}. We will come back to 
this point again later.

We have calculated $T_d(L)$ from the correlation function between two 
neighboring spins at the center of the strip:
\begin{eqnarray}
c_{L/2} = \langle s_{\frac{L}{2},j} s_{\frac{L}{2}+1,j}\rangle.
\label{defcL/2}
\end{eqnarray}
In the two phase coexistence region $c_{L/2}$ is large and positive 
since the two spins are preferably aligned.  If an interface is present, 
$c_{L/2}$ drops to smaller values, since in many configurations when the 
interface is located at the center of the strip the two spins tend to have 
opposite values.
We identify $T_d(L)$ as the maximum of the temperature derivative of 
$c_{L/2}$. There is obviously no sharp phase transition on an $L \times 
\infty$ system and in the present case $T_d(L)$ corresponds to a 
pseudocritical point.

\begin{figure}[b]
\centerline{
\rotate[r]{\psfig{file=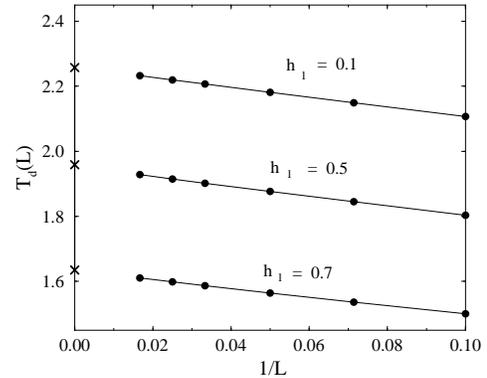,height=7cm}}}
\vskip 0.2truecm
\caption{Plot of $T_d(L)$ vs $1/L$ for $J=1$ and various values of the 
surface field $h_1$. The crosses on the vertical axis are the values of 
the wetting temperatures derived from Eq.(\ref{twexact}) where the data 
points are expected to scale to for $L \to \infty$.}
\label{FIG03}
\end{figure}

As $L \to \infty$, $T_d(L)$ scales as (\ref{wet}) towards the wetting 
temperature, with the two dimensional wetting exponent $\beta_s = 1$ 
\cite{abraham}.
Figure \ref{FIG03} shows a plot of $T_d(L)$ vs $1/L$ for various values 
of the surface field $h_1$. On the vertical axis the exact values of 
the wetting temperatures $T_w(h_1)$, derived from Eq. (\ref{twexact}), 
are shown. As can be seen from the figure the scaling behavior of 
$T_d(L)$ is in good agreement with Eq. (\ref{wet}).

The shift of $T_d(L)$ as $L \to \infty$ was also investigated in Monte Carlo 
simulations by Albano {\em et al.} \cite{albano}. They also found good 
agreement with (\ref{wet}).

Table I shows a comparison between the values of the wetting temperatures 
extrapolated from a finite size scaling of the DMRG data for $T_d(L)$ and 
the exact values given by Eq. (\ref{twexact}); the agreement 
is up to the fourth decimal digit. We stress that the relevant source of 
errors for $T_d(L)$ is in the calculation of the numerical derivative of 
$c_{L/2}$; DMRG calculations provide values of $c_{L/2}$ which are very
accurate.

\end{multicols} \widetext
\begin{table}[hb]
\caption{Comparison between DMRG data and exact values of the wetting 
temperatures $T_w$ for different values of the surface field $h_1$. 
The number between parenthesis are the error on the last two digits of 
the DMRG results.}
\label{T1}
\begin{tabular}{cdddddd}
$h_1$& 0.1 & 0.5 & 0.7 & 0.9 & 0.99  \\
\tableline
DMRG & 2.25761(77) & 1.95814(77) & 1.63532(77) & 
1.10753(77) & 0.58848(77) \\
Exact \cite{abraham} & 2.25710 & 1.95845 & 1.63496 & 
1.10745 & 0.58845 \\
\end{tabular}
\end{table}
\begin{multicols}{2} \narrowtext

\section{Effect of gravity}
\label{sec:phd}

It is convenient, before discussing the full phase diagram of the 
system described by the Hamiltonian (\ref{ham}) to focus on its 
ground state properties.
The two states with all spins positive or negative are degenerate, 
with an energy per spin equal to:
\begin{eqnarray}
\epsilon_{o} = - \left( 2-\frac{1}{L} \right) J .
\end{eqnarray}
If $g$ is large and positive the ground state configuration is:
\begin{eqnarray}
s_{i,j} = \left\{
\begin{array}{cl}
+1 & \,\,\, {\mbox if} \,\,\,  1 \leq i \leq \frac{L}{2} \\
   &       \\
-1 & \,\,\, {\mbox if} \,\,\,  \frac{L}{2}+1 \leq i \leq L ,
\end{array}
\right.
\label{groundg}
\end{eqnarray}
with an energy per spin equal to:
\begin{eqnarray}
\epsilon_{+-} = - \left( 2-\frac{3}{L} \right) J + \frac{2 h_1 }{L}
- \frac{g L}{2} .
\end{eqnarray}
Alternatively, if $g$ is large and negative the ground state has all spins
reversed with respect to the configuration (\ref{groundg}); in this 
case the energy is
\begin{eqnarray}
\epsilon_{-+} = - \left( 2-\frac{3}{L} \right) J - \frac{2 h_1 }{L} 
+ \frac{g L}{2} .
\end{eqnarray}

The ground state is thus double degenerate if $\epsilon_o < \epsilon_{+-}$
and $\epsilon_o < \epsilon_{-+}$, which yields the following condition:
\begin{eqnarray}
\frac{4(h_1-J)}{L^2} \,\,\, < \,\, g \,\, < \,\,\, \frac{4(h_1+J)}{L^2}
\label{scalg}
\end{eqnarray}
At $T=0$ the two phase coexistence region shrinks as $1/L^2$ towards the 
$g=0$ axis. The previous calculation is valid only for $h_1 < J$; if the 
surface field is larger than $J$ (this implies also $T_w = 0$) the range of 
values of $g$ for which one has phase coexistence at $T=0$ is given by:
\begin{eqnarray}
0 \,\,\, < \,\, g \,\, < \,\,\, \frac{4(h_1+J)}{L^2}.
\label{scalg0}
\end{eqnarray}
In the rest of the paper we will consider only the case $h_1 < J$.

\begin{figure}[b]
\centerline{
\rotate[r]{\psfig{file=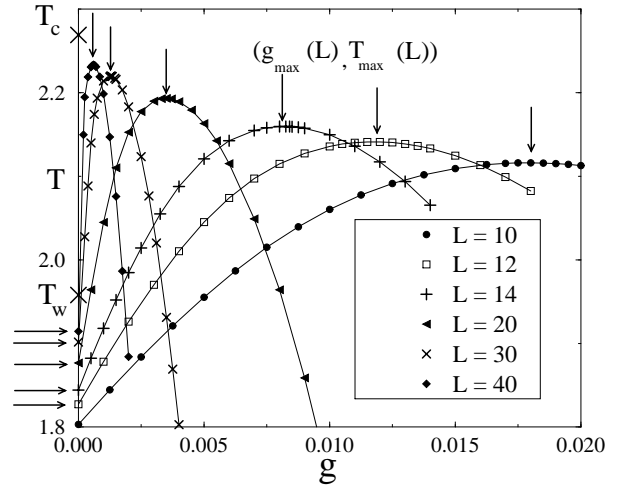,height=9cm}}}
\vskip 0.2truecm
\caption{Phase diagram of the model in the $(g,T)$ plane. The curves are
the phase boundaries between the two phase coexistence and the single phase
regions for different values of the strip width. Notice the scaling to the
wetting temperature (horizontal arrows) and to the critical point (vertical
arrows).}
\label{FIG04}
\end{figure}

The phase diagram of the model in a $(g,T)$ plane for $h_1=0.5$, $J=1$
is shown in Fig. \ref{FIG04}. The curves indicate the phase boundaries 
between the two phase coexistence region (area below the curves) and a 
single phase region (above) for different values of the strip width $L$
\cite{footnote1}. 
When continued down to $T=0$ the phase boundaries meet the horizontal 
axis at the values of $g$ equal to the extremes of the interval 
(\ref{scalg}).

The phase boundaries cross the axis $g=0$ at the interface delocalization 
temperatures $T_d(L)$ (indicated by horizontal arrows in Fig. \ref{FIG04}),
which scale to the wetting temperature $T_w$, indicated by a cross on the
$g=0$ axis of Fig. \ref{FIG04}.
The striking feature of the phase diagram \cite{jos} is that for non-zero 
gravity phase coexistence extends at higher temperatures with respect to 
the case $g=0$
due to the competing effect of surface and gravitational fields. At negative 
$g$ the two phase coexistence is suppressed at lower temperatures than for 
$g=0$, since in this case gravity enhances the effect of the surface fields.

The phase boundary maxima [$g_{\rm max}(L)$,$T_{\rm max}(L)$], indicated by 
the vertical arrows in Fig. \ref{FIG04}, shift towards the bulk critical 
point $T=T_c$, $g=0$ and can be identified as the finite system (pseudo)
critical points.
As $L \to \infty$ the whole critical region shifts towards the $g=0$ axis. 
This is in agreement with the results of Van Leeuwen and Sengers 
\cite{vanleeuwen} who pointed out that in presence of gravity there 
is no criticality in an infinitely extended system.
 
The phase diagram of Fig. \ref{FIG04} is in agreement with the mean field 
results of Rogiers and Indekeu \cite{jos}, who, in addition, found tricritical 
points located in the phase boundaries, which separate continuous from 
first order transition lines. Obviously these features are not found in 
two dimensions where the wetting transition is always critical, but they 
should be found in three dimensions where the wetting tricritical point 
is at $T > 0$.

Finite size scaling \cite{jos,fisher,nakanishi,vanleeuwen} predicts that 
the critical point shifts as follows:
\begin{eqnarray}
T_{\rm max}(L) - T_c \sim L^{-y_T},
\label{shiftTmax}
\end{eqnarray}
\begin{eqnarray}
g_{\rm max}(L) \sim L^{-(1+y_H)}.
\label{shiftgmax}
\end{eqnarray}
$y_T$ and $y_H$ are the thermal and magnetic exponents of the Ising model, 
which in two dimensions are $y_T=1$ and $y_H=15/8$.

\begin{table}[h]
\caption{The values of the extrapolated critical temperature from the scaling
analysis of $T_{\rm max}(L)$ for different values of the surface field $h_1$, 
when $h_1 < J$. The exact value is $T_c = 2.269185\ldots$}
\label{T2}
\begin{tabular}{cddd}
$h_1$& 0.1 & 0.5 & 0.99 \\
\tableline
DMRG & 2.269(3) & 2.272(3) & 2.271(3)
\end{tabular}
\end{table}

In deriving Eq. (\ref{shiftgmax}) one assumes, as done by van Leeuwen and 
Sengers \cite{vanleeuwen}, that the gravitational constant $g$ times a 
length scales as a bulk constant field. This relates the scaling of $g$
to the bulk magnetic exponent $y_H$.

\begin{figure}[b]
\centerline{
\rotate[r]{\psfig{file=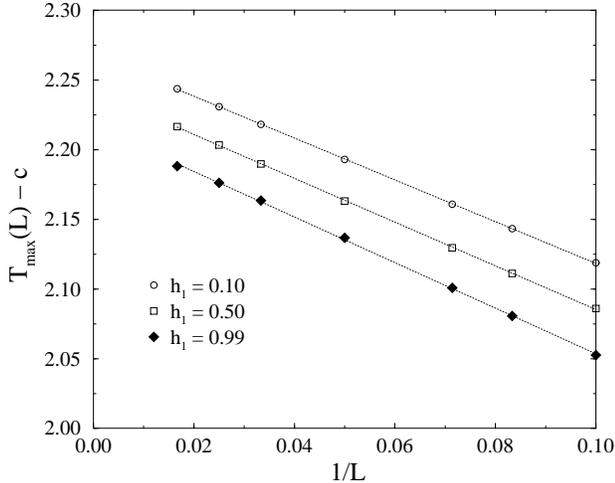,height=9cm}}}
\vskip 0.2truecm
\caption{Scaling of $T_{\rm max}(L) - c$ vs $1/L$ for different surface 
fields.  To avoid overlapping of data the values of $T_{\rm max}(L)$ have 
been shifted by a constant $c$, with $c=0$ for $h_1 = 0.1$, $c=0.3$ for 
$h_1 = 0.5$ and $c = 0.6$ for $h_1 = 0.99$. Dotted lines are linear 
fits of the data points.}
\label{FIG05}
\end{figure}

The DMRG data for $T_{\rm max}(L)$ are in very good agreement with the 
scaling relation (\ref{shiftTmax}), as shown in Fig. \ref{FIG05}.
Table \ref{T2} shows the values of $T_c$ extrapolated from the finite size
scaling analysis of $T_{\rm max}(L)$ by means of iterated fits. Results are
in good agreement with the exact value. 

\begin{figure}[b]
\centerline{
\rotate[r]{\psfig{file=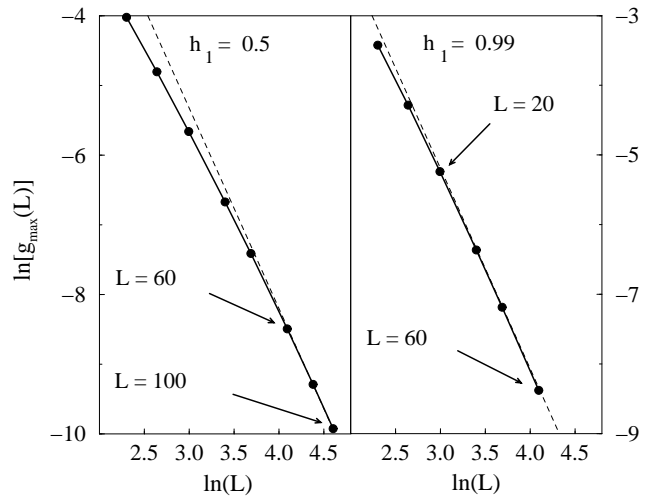,height=9cm}}}
\vskip 0.2truecm
\caption{Plot of $\ln[g_{\rm max}(L)]$ vs $\ln L$ for $h_1 = 0.5$ (left)
and $h_1 = 0.99$ (right). Error bars are smaller than symbol sizes. The 
dashed lines correspond to a slope $-2.875$ and are drawn as a guide to 
the eye.}
\label{FIG06}
\end{figure}

The finite size scaling along the gravitational field direction is more 
intriguing. Figure \ref{FIG06} shows a plot of $\ln[g_{\rm max}(L)]$ vs 
$\ln(L)$ for $h_1 = 0.5$ and $h_1 = 0.99$. In both cases there is 
agreement with the scaling relation (\ref{shiftgmax}): as $L$ grows the 
data points approach the dashed lines which have slope $-2.875$ (the 
value of the exponent $-1-y_H$ for the two dimensional Ising model).
Notice that for $h_1 = 0.99$ the asymptotic behavior sets in already for
$L \geq L_0 \approx 20$ while for $h_1 = 0.5$ this happens only for 
$L \geq L_0 \approx 60$.

\begin{figure}[b]
\centerline{
\rotate[r]{\psfig{file=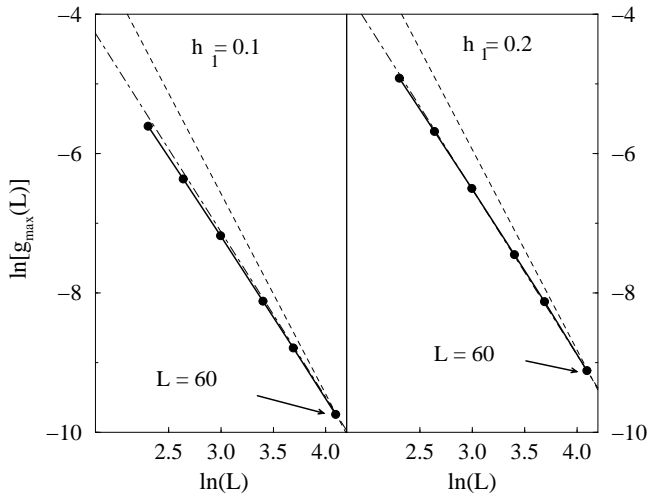,height=9cm}}}
\vskip 0.2truecm
\caption{As in Fig. \ref{FIG06} for $h_1 = 0.1$ (left) and $h_1 = 0.2$
(right). Dashed and dashed-dotted lines have slopes $-2.875$ and $-2.375$
respectively.}
\label{FIG07}
\end{figure}

The scaling behavior of $g_{\rm max}(L)$ for smaller surface fields is shown
in Fig. \ref{FIG07}; in this case the deviation from the expected exponents 
is so strong that one concludes that either (\ref{shiftgmax}) does not hold 
for small $h_1$ or the asymptotic behavior sets in only for strip widths 
much larger than those analyzed.

Following Fisher and Nakanishi \cite{fisher,nakanishi} who investigated 
the critical point shift in an Ising model confined between identical 
walls, one expects a scaling of $g_{\rm max}(L)$ of the type:
\begin{eqnarray}
g_{\rm max}(L) = L^{-(1+y_H)} \,\,\,\Omega\left(h_1 L^{\Delta_1 y_T} \right),
\label{scalansatz}
\end{eqnarray}
where $\Omega(x)$ is a scaling function and $\Delta_1$ the surface gap 
exponent (recall that $\Delta_1 = 1/2$ for the two-dimensional Ising
model). 
The fact that the surface field enters in the form of a scaling variable 
$h_1 L^{\Delta_1 y_T}$ is a direct consequence of the scaling form of 
the singular part of the surface free energy \cite{fisher,nakanishi}.

For $L \to \infty$ one should recover from (\ref{scalansatz}) the scaling 
relation (\ref{shiftgmax}), which implies that:
\begin{eqnarray}
\lim_{x \to \infty} \Omega(x) = \Omega_{\infty} \neq 0
\end{eqnarray}
When $h_1$ is sufficiently large and for not too small values of $L$, 
the scaling function in (\ref{scalansatz}) can be replaced by its 
asymptotic value $\Omega_{\infty}$. In this limit, which we refer to 
as the saturated regime, $g_{\rm max}(L)$ becomes practically independent 
of $h_1$.

From the analysis of the numerical data for $h_1 = 0.99$ one finds that 
$\Omega(x)$ saturates for $x > x_0 \approx \sqrt{20}$. Consequently, for
$h_1 = 0.5$, $0.2$ and $0.1$ the saturated regime is expected to occur
for $L > L_0 \approx 80$, $L > L_0 \approx 500$ and $L > L_0 \approx 2000$
respectively.
Notice that a saturation value of $L_0 \approx 80$ for $h_1 = 0.5$ is in 
agreement with our numerical data. Strips of width $L = 500$ or $L = 2000$ 
are beyond the possibilities of our numerical investigation; actually,
calculations for $L > 100$ (the largest size analyzed in the present 
work) are feasible, but for such large systems the value of $g_{\rm max}(L)$
is very small and affected by large relative error bars that make the scaling 
analysis difficult.

For $h_1 = 0$ one has $g_{\rm max}(L) = 0$ since the phase boundaries of 
Fig. \ref{FIG04} become symmetric with respect to the $g=0$ axis. For very 
small surface fields one expects that $g_{\rm max}(L)$ scales linearly with 
$h_1$ \cite{footnote2}; this implies that:
\begin{eqnarray}
\Omega(x) \sim x \hspace{1cm} {\mbox {\rm for}}\,\,\, x \to 0.
\end{eqnarray}

Therefore, in the strongly ``undersaturated" regime, where $x = h_1 
L^{\Delta_1 y_T} \ll \sqrt{20}$, one expects a scaling 
behavior in $L$ of the type:
\begin{eqnarray}
g_{\rm max}(L) \sim L^{-(1 + y_H - \Delta_1 y_T)}
\label{undersat}
\end{eqnarray}
In Fig. \ref{FIG07} the dashed-dotted lines have slopes equal to $-2.375$, 
which is the value of the exponent $-1 - y_H + \Delta_1 y_T$ for the two 
dimensional Ising model. As can be seen from the figure the numerical data 
are in good agreement with a scaling of type (\ref{undersat}).

\begin{figure}[b]
\centerline{
\rotate[r]{\psfig{file=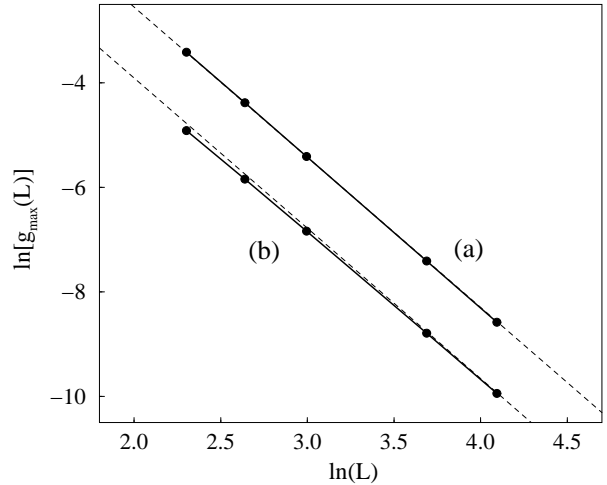,height=9cm}}}
\vskip 0.2truecm
\caption{
Plot of $\ln [g_{\rm max} (L)]$ vs $\ln L$ at constant values of 
$x = h_1 L^{\Delta_1 y_T}$. The data correspond to (a) $x = {\sqrt{10}}$ 
and (b) $x = {\sqrt{0.4}}$. Dashed lines have slope ${-2.875}$.
}
\label{FIG08}
\end{figure}

We stress that (\ref{undersat}) is not the asymptotic behavior of 
$g_{\rm max}(L)$ as $L \to \infty$, since for sufficiently large $L$ the
scaling is of type (\ref{shiftgmax}). To verify this we have calculated 
$g_{\rm max}(L)$ for different strip widths at constant values of
$x = h_1 L^{\Delta_1 y_T}$. The data points are shown in Fig. \ref{FIG08} 
and they are calculated in the undersaturated regime ($x < x_0 \approx 
\sqrt{20}$); the agreement with the exponent $1+y_H = 2.875$ is very good.

To conclude this section we point out that also for the temperature
one expects a shift of the type \cite{fisher,nakanishi}:
\begin{eqnarray}
T_c - T_{\rm max}(L) = L^{-y_T} \,\, \Gamma( h_1 L^{\Delta_1 y_T}).
\end{eqnarray}
Nakanishi and Fisher \cite{nakanishi} analyzed the behavior of $\Gamma(x)$ 
in a mean field model confined between identical walls and found that 
$\Gamma(x)$ depends very weakly on its argument $x$. This is also found
in the present study, as it can be seen from the fact that
(1) the slopes of the data points of Fig. \ref{FIG05} are almost equal,
i.e. they do not depend sensibly on the value of the surface field, and
(2) the points are very well fitted by straight lines.

\section{Magnetization Profiles}
\label{sec:pro}

\begin{figure}[b]
\centerline{
\rotate[r]{\psfig{file=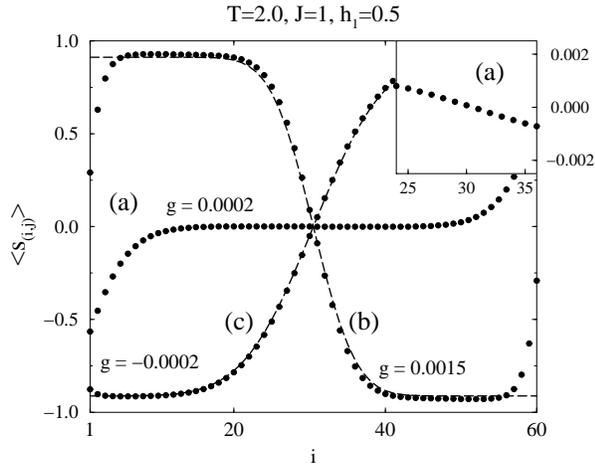,height=8cm}}}
\vskip 0.2truecm
\caption{Magnetization profiles at a fixed temperature $T=2.0$ and for
three different values of the gravitational constants calculated by DMRG
(filled circles). The profile (a) is in two phase coexistence region.
(b) and (c) are profiles in the single phase region situated to the left
and to the right of the two phase coexistence region of Fig. \ref{FIG04}
respectively. 
The dashed lines are the result of the capillary wave theory calculation
as given by (\protect\ref{sosmag}).
Inset: enlargement of the profile (a) in the center of the strip.}
\label{FIG09}
\end{figure}

\begin{figure}[b]
\centerline{
\rotate[r]{\psfig{file=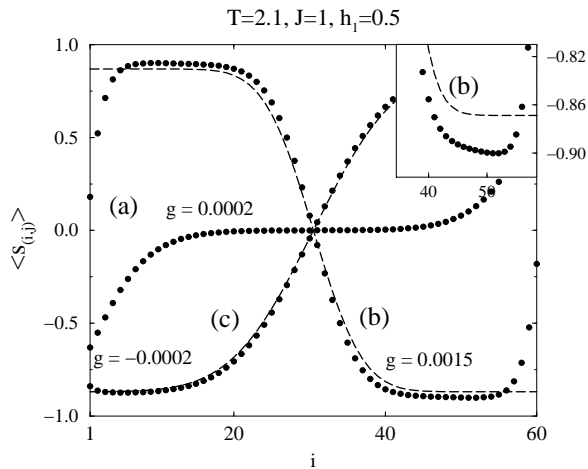,height=8cm}}}
\vskip 0.2truecm
\caption{As in Fig. \ref{FIG09} at higher temperature ($T=2.1$).
Notice that the discrepancy between the profiles in the interfacial region 
((b) and (c)) and those predicted by the capillary wave theory increase
at higher temperatures.
Inset: enlargement of the profile (b) in the vicinity of one wall. }
\label{FIG10}
\end{figure}

Figures \ref{FIG09} and \ref{FIG10} show some examples of magnetization 
profiles calculated by DMRG for various values of the gravitational 
constant and temperatures. The profiles (a) refer to points below the 
phase boundaries of Fig. \ref{FIG04}, i.e. in the two phase coexistence 
region. The two coexisting phases are expected to have magnetization 
profiles similar to those depicted in Fig. \ref{FIG02}(a); the profiles (a)
are actually the result of the average over the two phases.

The magnetization in the two phase coexistence region does not decay to a 
bulk constant value far from the walls as in Fig. \ref{FIG02}(a). The 
inset of Fig. \ref{FIG09} shows an enlargement of the profile (a) at the 
center of the strip: due to the presence of gravity the profile follows
a straight slightly inclined line. We stress that the accuracy of the DMRG
results for the magnetization profiles is very high, so error bars are much 
smaller than symbol sizes also in the scale of the inset of Fig. \ref{FIG09}.

In the single phase region (Figs. \ref{FIG09},\ref{FIG10} (b,c)) the system
develops an interface: gravity has the tendency to localize the interface 
at the center of the strip in a configuration of minimal energy. The larger the
value of $|g|$ the stronger the localization effect, as it can be seen in the
figures: the profiles (b) correspond to a value of the gravitational constant 
$7.5$ times larger than the profiles (c). Notice also in the profiles (b) the 
competing effect of surface fields and gravity which is visible in the vicinity 
of the walls.

The dashed lines of Figs. \ref{FIG09}-\ref{FIG10} are the magnetization
profiles predicted by the capillary wave theory, which are derived in the
Appendix and are given by:
\begin{eqnarray}
m(l) = \frac{2 m_0}{\sqrt{\pi}}  \int_0^{l/\xi_\perp} dt \, e^{-t^2},
\label{sosmag}
\end{eqnarray}
Here $2 \xi_\perp$ is the average interfacial width and $m_0$ is the bulk 
magnetization of the Ising model in absence of gravity. All the parameters 
appearing in (\ref{sosmag}) are known exactly, and the dashed lines of 
Fig. \ref{FIG09}, \ref{FIG10} are not the results of a fitting. 

Capillary wave theory profiles agree very well with the DMRG results 
especially at low temperatures where the approximations introduced are 
very good. Eq. (\ref{sosmag}) is valid in the limit $\xi_\perp \ll L$ where
the effects of the walls can be neglected. This condition is satisfied
at low temperatures and at not too small values of $|g|$. Notice also that
if $|g|$ is too large the magnetization profile (\ref{sosmag}) far from 
the interfacial region differs sensibly from the DMRG results, since the
effect of gravity in that region has been neglected. This can be seen 
more clearly in the inset of Fig. \ref{FIG10}.

\section{Conclusions}
\label{sec:con}

In this article we have studied the critical behavior of an Ising model
confined between opposing walls and subject to a bulk ``gravitational"
field. The competing effects of surface and bulk fields restore phase 
coexistence up to the bulk critical temperature, in agreement with the
results of a mean field study of the model \cite{jos}.
The strong thermal fluctuations in two dimensions do not affect the mean
field results, which {\em a fortiori} should also be valid in three 
dimensions where fluctuation effects are weaker.

Wetting plays an important role in the model as it was found in the
studies in the absence of gravity \cite{parry,parryA,broch}. However, 
limiting the analysis to $g=0$ causes us to miss much of the interesting 
physics that arises when gravity is included.
In particular, one misses the critical point of the confined system \cite{jos}, 
which we have identified with [$g_{\rm max}(L)$,$T_{\rm max}(L)$], the 
maximum of the phase boundaries separating the two-phase coexistence from
the single phase regions.

We have performed a detailed analysis of the critical point shift as 
$L \to \infty$ and found that temperature and gravitational constant 
scale in agreement with previous finite size scaling hypothesis 
\cite{jos,fisher,nakanishi,vanleeuwen}.
Along the gravitational field direction in the limit of small surface 
fields a crossover behavior between two different scaling regimes is 
found.

This limit has been considered recently in studies of critical adsorption 
\cite{carl,ciach}. Desai {\em et al.} \cite{carl} in a experiment on a
binary liquid mixture were able to investigate the weak surface field
regime by chemically modifying the surface of the solid substrate which
is in contact with the liquid.
They found an unexpected behavior of the adsorption for small surface fields,
in possible disagreement with scaling theory. This issue was discussed recently 
also by Ciach {\em et al.} \cite{ciach}.
In general we expect interesting physics to arise in the limit $h_1 \to 0$
due to the interplay between bulk and surface criticality, as we found in the 
analysis of the scaling behavior of the gravitational constant for the model 
studied in this article.

{\bf Acknowledgments} - It is a pleasure to thank C.J. Boulter, R. Dekeyser, 
M.S.L. du Croo de Jongh, C. Franck, J.O. Indekeu, J.M.J. van Leeuwen, 
A.O. Parry and J. Rogiers for stimulating discussions. 
We acknowledge also fruitful correspondence with A. Macio\l ek and J. Stecki. 
E.C. would like to thank J. Sznajd for the kind hospitality during his visit 
to the Institute for Low Temperature and Structure Research of the Polish 
Academy of Sciences (Wroc\l aw) where part of this work was done. 
A.D. is grateful to R. Dekeyser for financial support during his visit at 
the Institute for Theoretical Physics of the Catholic University of Leuven.
E.C. is financially supported by KULeuven Research Fund F/96/20.

\section*{Appendix: Capillary-wave theory}
\label{sec:sos}

\begin{figure}[b]
\centerline{
\rotate[r]{\psfig{file=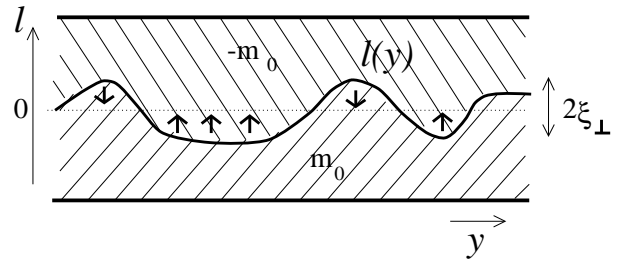,height=8cm}}}
\vskip 0.2truecm
\caption{Example of an interfacial configuration described by a 
continuous single-valued function $l(y)$ as in the solid-on-solid 
approximation. All the spins at the two sides of the interface are 
fixed and equal to $\pm m_0$. The arrows denote the elastic force 
that tends to bring the interface in its equilibrium position.}
\label{FIG11}
\end{figure}

In capillary wave theory \cite{widom} one assumes that a sharp interface
separates two regions of constant magnetization. We take magnetizations 
equal to $\pm m_0$, the bulk magnetization of the Ising model in absence of 
external fields, which is known exactly. 
In this approximation gravity does not affect the magnetization far from the 
interfacial region and bulk fluctuations are neglected. In the continuum limit 
the interface is described by a single valued function $l(y)$ where $y$ is the 
coordinate along the wall ($-\infty < y < \infty$) and $l$ denotes the 
displacement of the interface from the center of the strip (see Fig. \ref{FIG11}).
This is a solid-on-solid (SOS) approximation where overhangs are neglected 
\cite{stecki}.

The continuum Hamiltonian is given by:
\begin{eqnarray}
H\left[l(y)\right] &=& \int_{-\infty}^{+\infty} dy \,\, 
\left\{ \frac{\sigma _0}{2} 
\left(\frac{dl}{dy}\right)^2 + U(l) \right\}
\label{intham}
\end{eqnarray}
where $\sigma_0$ is the surface tension and $U(l)$ the potential acting on the
interface.
The partition function is given by:
\begin{eqnarray}
Z = \int {\cal{D}}l(y) \,\,e^{- \beta H[l(y)]} 
\label{partfun}
\end{eqnarray}
where $\beta$ denotes the inverse temperature. In (\ref{partfun})
we integrate over all the possible interface shapes described by
single valued functions. Due to well-known relations between path 
integrals and quantum mechanics \cite{lipowski}, the previous problem 
can be mapped into a one dimensional quantum problem which consists 
in solving the following Schr\"odinger equation:
\begin{eqnarray}
\left(-\frac{1}{2 \sigma_0 \beta^2} \frac{d^2}{dl^2} + U(l)
\right) \psi_n(l) = E_n \psi_n(l)
\label{schroed}
\end{eqnarray}
The ground state wave function squared $|\psi_0(l)|^2$ denotes the
probability of finding the interface at a position $l$.
The potential has the following form:
\begin{eqnarray}
U(l;L) = W(l;L) + V_g(l)
\end{eqnarray}
Here $W(l;L)$ is the confining potential due to the presence of the walls:
\begin{eqnarray}
W(l;L) = \left\{\begin{array}{cc}
0\,\,\, & \mbox{if}\,\, |l| \leq L/2 \\
 & \\
+\infty \,\,\,& \mbox{if} \,\, |l|>L/2 
\end{array}
\right.
\label{confining}
\end{eqnarray}
$V_g(l)$ represents the contribution of gravity to the interfacial potential
and can be calculated from the microscopic Hamiltonian (\ref{ham}).
If the interface is located between the k-th and (k+1)-th spin of the 
system gravity gives the following contribution to the energy:
\begin{eqnarray}
Q(k) &=& g m_0 \left\{ \sum_{i=1}^k (2i - 1 - L) - 
\sum_{i=k+1}^L (2i - 1 - L) \right\}  \nonumber\\
&=& 2 m_0 g k (k-L)
\end{eqnarray}
Shifting appropriately the origin of the coordinates one finds:
\begin{eqnarray}
V_g(l) &=& 2 m_0 gl^2 + c
\end{eqnarray}
where $c$ is a constant.

Neglecting the effect of the confining potential (\ref{confining}) becomes 
the Schr\"odinger equation for a harmonic oscillator, with ground state 
wave function equal to:
\begin{eqnarray}
\psi_0(l) &=& \frac{e^{-l^2/(2 \xi_\perp^2)}}{\pi^{1/4} \sqrt{\xi_\perp}}.
\label{harmground}
\end{eqnarray}
$\xi_\perp$ denotes the interfacial width:
\begin{eqnarray}
\xi_\perp= \sqrt{\frac{T}{2 \sqrt{|g| m_0 \sigma_0}}}.
\end{eqnarray}
Here $m_0$ is the bulk magnetization and $\sigma_0$ is the surface tension
which are known exactly for the $d=2$ Ising model at $g=0$. 
The analysis is valid for $\xi_\perp \ll L$ where the probability of finding
the particle outside the walls is negligible and the confining potential
$W(l;L)$ can be ignored.

From (\ref{harmground}) one can easily calculate the magnetization profile 
as function of the distance from the center of the strip $l$:
\begin{eqnarray}
m(l) = m_0 \left\{ \int_{-\infty}^l ds\,\, \left| \psi_0(s) \right|^2
- \int_l^{+\infty} ds\,\, \left| \psi_0(s) \right|^2
\right\} 
\end{eqnarray}
which yields the result given in (\ref{sosmag}).

\end{multicols}


\begin{references}
\bibitem{footnote1} The phase boundaries have been calculated with the same
method that has been used for the calculation of $T_d(L)$: at fixed $g$ we
identified the phase boundaries with the maximum of the temperature derivative
of $c_{L/2}$.
\bibitem{footnote2} Fisher and Nakanishi \cite{fisher} argued that it is
plausible to expect a linear scaling of the critical bulk field as function
of $h_1$, for small $h_1$ in the case of confinement between identical walls.
\end{references}
\end{document}